\documentclass[aps,prb,twocolumn,superscriptaddress]{revtex4-1}

\usepackage{hyperref} 
\usepackage{graphicx}
\usepackage{amsmath}
\usepackage{amssymb}
\usepackage{float}

\DeclareGraphicsExtensions{.png,.jpg,.eps}
\usepackage{xcolor}

\def\bk{{\bold{k}}}
\def\br{{\bold{r}}}

\def\m1{{^{-1}}}

\begin{document}

\author{Aline Ramires}
\email[]{aline.ramires@ictp-saifr.org}
\affiliation{ICTP-SAIFR, International Centre for Theoretical Physics - South American Institute for Fundamental Research, S\~{a}o Paulo, SP, 01140-070, Brazil}
\affiliation{Instituto de F\'{i}sica Te\'{o}rica - Universidade Estadual Paulista, S\~{a}o Paulo, SP, 01140-070, Brazil}

\author{Jose L. Lado}
\affiliation{Institute for Theoretical Physics, ETH Zurich, 8093 Zurich, Switzerland}

\title{
	Impurity-induced triple point fermions in twisted bilayer graphene
}

\begin{abstract}
Triple point fermions are elusive electronic excitations that generalize Dirac
	and Weyl modes beyond the conventional high energy paradigm. Yet,
	finding real materials naturally hosting these excitations at the Fermi
	energy has remained challenging. Here we
	show that twisted bilayer
	graphene is a versatile platform to realize robust triple point
	fermions in two dimensions. 
	In particular, we establish that the introduction of localized impurities lifts one of the two degenerate Dirac cones, yielding triple
	point fermions at charge neutrality. Furthermore, we show that the
	valley polarization is preserved for certain impurity locations in the
	moire supercell for both weak and strong impurity potentials. 
	We finally show that in the presence of interactions, a symmetry
	broken state with local magnetization
	can develop out of the triple point bands, which can be selectively
	controlled by electrostatic gating.
	Our results put forward twisted bilayer graphene as a
	simple solid-state platform to realize triple point fermions at charge
	neutrality, and demonstrate the non-trivial role of impurities in moire
	systems.
\end{abstract}

\date{\today}

\maketitle


\section{Introduction}

Topological semimetals have attracted a lot of attention in the past years, as
they provide solid-state platforms to realize analogs of relativistic
particles \cite{Yan2017,Burkov2016}, namely Dirac and Weyl fermions
\cite{Armitage2018}, whose spinorial form stems from Lorentz invariance.
However, space group symmetries in materials
provide an even more versatile playground,
as they impose only a subset of the symmetries inflicted by Lorentz invariance, 
enabling novel types of effective particles to emerge
beyond the conventional high energy paradigm
 \cite{ PhysRevX.6.031003, Cheung2018,PhysRevB.95.235116, Bradlyn2017,
PhysRevLett.116.186402,Bzdusek2016,PhysRevLett.116.186402,Wang2016,
Soluyanov2015,PhysRevB.96.045121,Ma2018,Lv2017,PhysRevX.8.041026,
PhysRevB.95.205201,Owerre2017,
PhysRevLett.122.076402,PhysRevX.7.041069}. 
Among these possibilities,
triple point fermions
 \cite{PhysRevX.6.031003,
Cheung2018,PhysRevLett.119.136401,Shekhar2017,Gao2018,
PhysRevLett.119.256402,PhysRevB.96.241204,PhysRevB.98.045134,
PhysRevLett.121.176601,PhysRevB.98.155122,PhysRevMaterials.2.081201} 
are exotic excitations displaying unusual magneto-transport
phenomena including large negative magneto-resistance and helical
anomaly \cite{WengPRB2016, PhysRevB.95.195165, Chang2017}, in constrast to the chiral anomaly observed in Weyl semimetals \cite{Armitage2018}. 
From the material science point of view,
recent proposals suggest the presence of triple point fermions
away from the Fermi energy in Heusler compounds
 \cite{PhysRevLett.119.136401,PhysRevB.99.045144}. 
Experimentally,  angle-resolved
photoemission spectroscopy (ARPES) measurements
have observed triple point fermions \cite{Lv2017,
Ma2018,PhysRevLett.122.076402}, but only weak signatures of the expected exotic transport phenomena
could be observed given the distance of the triple points to the Fermi level
 \cite{Gao2018,PhysRevB.95.195165}. Thus, materials displaying robust
 triple point
fermions at the Fermi energy have remained elusive,
frustrating the experimental exploration of their associated exotic properties.

\begin{figure}[t!]
\centering
    \includegraphics[width=\columnwidth]{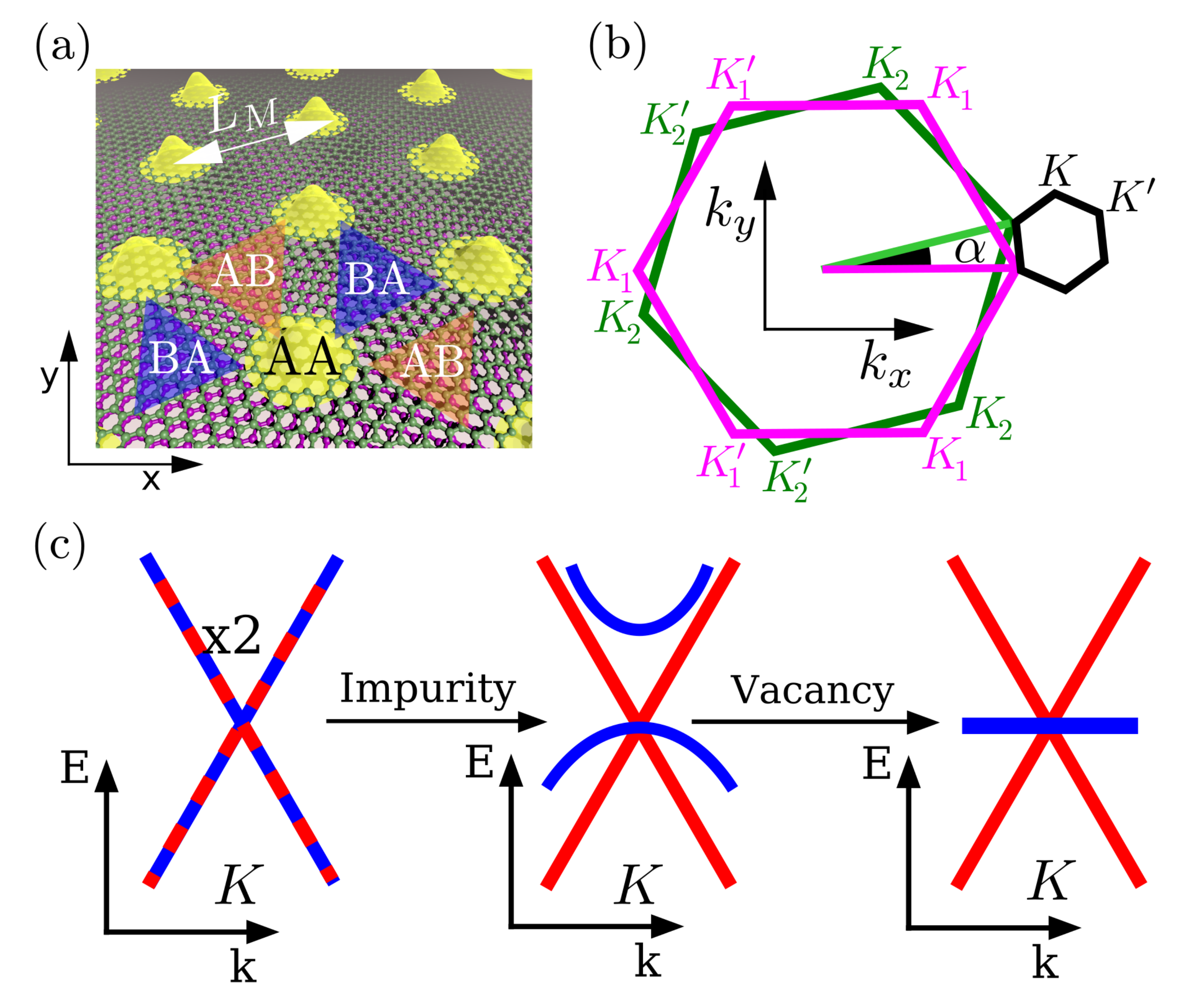}

\caption{
	(a) Perspective view of twisted bilayer graphene, highlighting the electronic density concentrated in the AA regions (yellow blobs), the AB/BA regions (orange/blue triangles), and the moire length $L_M$.
	(b) Brillouin zones for the first (pink) and second (green) layer, and the emergent moire-Brillouin zone (black), with the respective $K$ and $K'$ points.
	(c) Sketch of the effects of a weak impurity and a vacancy on the degenerate low energy Dirac cones, with the generation of a triple point at the Fermi level.
}
\label{fig1}
\end{figure}

Graphene is known for being an extremely clean platform to explore Dirac fermion
phenomena, with two spin degenerate Dirac cones at each $K$ point
 \cite{DiracMat}. Engineering triple point fermions out of graphene by lifting
the degeneracy of the Dirac points would require eliminating key symmetries,
which usually leads to a shift of the cones or to a gapped spectrum.
Introducing an additional level of complexity,
graphene multilayers have the potential to enlarge the degeneracy of the
Dirac cones, providing new routes for degeneracy lifting. 
Among them, twisted bilayer
graphene (TBG) (Fig. \ref{fig1}(a)) is an
especially promising candidate for displaying a band
structure with four-fold
degenerate Dirac cones in the reduced Brillouin zone (Fig. \ref{fig1}(b) and (c)) \cite{PhysRevB.86.155449,
Koshino2018, Kang2018},
providing a potential direction for the engineering
of triple points by the controlled reduction of symmetries. 

Here we show that impurities in TBG create robust triple point fermions. This unexpected feature stems from the interplay of the Dirac point degeneracy and the local nature of the impurity, which gives rise to mass generation in one of the Dirac cones, creating a triple point at charge neutrality independently of the strength and location of the impurity. Furthermore, we show that the valley polarization of the triple point can be controlled by the location of the impurities in the moire supercell. 
The manuscript is organized as follows: in Sec. \ref{sec:model} we introduce
the tight binding model
for twisted bilayer graphene, as well as a procedure to compute the expectation value
of the valley operator in real space. In Sec. \ref{sec:weak}, we numerically show the presence of triple points for weak impurities and provide a  low energy effective model which accounts for the triple point formation and discuss its robustness. We examine the vacancy limit, in Sec.\ref{sec:vac}. In Sec. \ref{sec:int}, we evaluate the effect of interactions
in this triple point system, showing that electronic doping allows one to selectively control the symmetry broken phases. Finally, in Sec. \ref{sec:con} we summarize our results and conclusions.

\section{Tight binding model for twisted bilayer graphene with local impurities}
\label{sec:model}

Here we focus on TBG superlattices with
long moire wavelength, i.e. small twisting angles
$\alpha$ (see Fig. \ref{fig1} (a)), but we would like to highlight that the presence of the triple point is independent of the magnitude of the twist angle, as discussed in detail in Appendix \ref{sec:largealpha}. In the small angle regime, the low energy model consists of two sets of honeycomb-like bands with strongly renormalized Fermi velocity, which vanishes at the magic angle
$\alpha \approx 1^\circ$ \cite{SantosPRL2007,bistritzer2011moire}. We model TBG
in the presence of impurities by the
real space Hamiltonian of the form:
\begin{eqnarray}
	\mathcal{H} = \mathcal{H}_0 + \mathcal{W},
	\label{hamil}
\end{eqnarray}
where $\mathcal{H}_0$  encodes the pristine
TBG tight binding Hamiltonian \cite{SboPRB2015}:
\begin{eqnarray}
	\mathcal{H}_0 = t \sum_{\langle i, j \rangle} c_i^\dagger c_j +
	\sum_{i,j} \bar t_{\perp} (\br_i,\br_j) c_i^\dagger c_j, 
\end{eqnarray}
and $\mathcal{W}$ describes a local impurity at site $n$
\begin{equation}
	\mathcal{W} = w c^\dagger_n c_n,
\end{equation}
where $c^\dagger_i$ ($c_i$) is the fermionic creation (annihilation) operator
at site $i$, $w$ is the impurity potential strength, $t$ is the nearest
neighbor hopping, $\langle i, j \rangle$ indicates the sum over first
neighbors, and $\bar t_{\perp} (\br_i,\br_j)$ is the distance-dependent interlayer
coupling taking a maximum value $t_\perp$ for perfect stacking
\footnote{
We take 
$
\bar t_{\perp}(\bold r_i,\bold r_j) = 
t_{\perp}
\frac{(z_i - z_j)^2 }{|\bold r_i - \bold r_j|^2}
e^{-\beta (|\bold r_i - \bold r_j|-d)},
$
where for simplicity we take
$d=3a$ the interlayer distance,
$\beta=3/a$ with $a$ the carbon-carbon distance.
}. As a reference, the values of the parameters in graphene are $t\approx$ 3 eV and $t_\perp \approx 300$ meV \cite{McCannRMP2013}.  The previous Hamiltonian is defined in a moire unit cell with $N = 4(3m^2_0 + 3m_0 + 1)$ sites, with $m_0$ an integer, and the magic angle regime is reached for $t_\perp/(tm_0) \approx 0.025$. For the sake of simplicity here we omit the spin degree of freedom. It should be understood that for the spinful scenario the triple-point degeneracy is in fact six-fold.

\begin{figure}[t!]
\centering
    \includegraphics[width=\columnwidth]{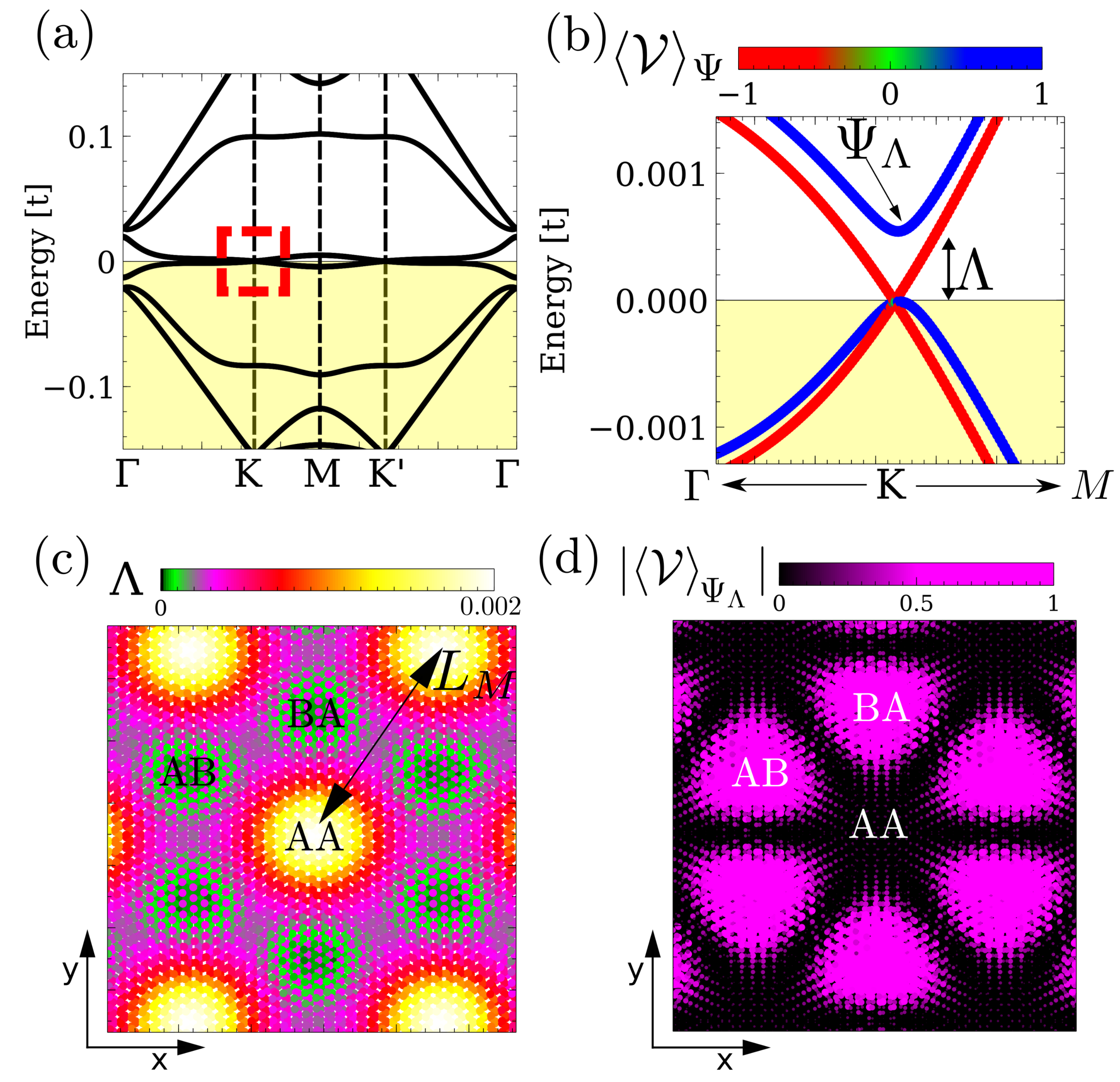}

	\caption{ 
(a) Band structure of twisted bilayer graphene with
	$\alpha=1.5^\circ$ in the presence of a weak impurity
	in the AB region.
	(b) A zoom at the $K$ point,
	showing the valley polarization $\langle \mathcal{V} \rangle_\Psi$
	of each state (color gradient)
	and the emergence of a gap of magnitude $\Lambda$.
(c) Map of the splitting $\Lambda$ as a function of the impurity position.
(d) Map of the valley polarization of $\Psi_\Lambda$ at the K point, as a function of the
	impurity position.  Here we took $m_0=11$, $t_\perp=0.3t$ and $w=0.5t$.
}
\label{fig2}
\end{figure}

In the absence of impurities, $w = 0$, the low energy spectra
of the previous Hamiltonian
consists of two Dirac cones at $K$ and
other two at $K'$ \cite{SantosPRL2007,bistritzer2011moire,
Kang2018,Koshino2018,Po2018}. 
This degeneracy can be understood by
the folding of the cone at $K_1$ from layer 1
and $K'_2$ from layer 2 at the same $K$ point in the moire Brillouin zone, as
can be seen from Fig. \ref{fig1} (b). Due to the approximate valley symmetry of the low energy graphene Hamiltonian,
the states associated with each decoupled layer $\ell$ in TBG can be labeled by
their valley number $\mathcal{V}_{\ell}$. In a real
space tight-binding formalism, such valley flavor can be computed using
the valley operator:
\begin{eqnarray}
	\mathcal{V}_\mathcal{\ell} =\frac{i}{3\sqrt{3}}
	\sum_{\langle\langle i,j \rangle\rangle \in \ell}
\eta_{ij} \sigma_z^{ij} c^\dagger_i c_j,
\end{eqnarray}
where $\langle\langle i,j \rangle\rangle$ denotes second neighbor sites, $\eta_{ij}=\pm 1$ is for clockwise or anticlockwise
hopping, and $\sigma^{ij}_z$ is a Pauli matrix associated with the sublattice
degree of freedom \cite{Ramires2018,PhysRevLett.120.086603}. 
 The interlayer hopping couples opposite valleys between the two layers, giving
 rise to a new quantum number  $\mathcal{V} = \mathcal{V}_1 - \mathcal{V}_2$,
 which is conserved in the absence of impurities. The addition of impurities
 generally introduces
inter-valley scattering, and therefore it is interesting to track the valley
polarization of the states. 




\section{Weak impurity limit}
\label{sec:weak}

We first consider the case of a weak impurity in twisted
bilayer graphene.
When the local impurity potential is turned on, $w \ne 0$, the four-fold
degeneracy of the states at $K$ and $K'$ is lifted, giving rise to triple
points at the Fermi level, as shown in Fig. \ref{fig2} (a) and (b), 
independently of
the position of the impurity in the moire pattern.
The location of the impurity controls the splitting
$\Lambda$
between the triple point and the higher
lying state $\Psi_\Lambda$. In particular, 
Fig. \ref{fig2} (c) 
shows that impurities located in the AA regions
create larger splittings. This can be understood from the larger amplitude of
the wave functions in these regions \cite{SantosPRL2007,bistritzer2011moire,
Kang2018,Koshino2018,Po2018,PhysRevLett.119.107201}, 
as schematically shown in Fig. \ref{fig1}(a). 
Interestingly, although a local impurity would be
expected to give rise to inter-valley mixing between the degenerate Dirac
points, we observe that the states associated with the triple point and $\Psi_\Lambda$ remain valley polarized for
impurities located in the AB/BA regions, as shown in Fig.
\ref{fig2} (d). 

\subsection{Low energy effective model}
\label{sec:lowem}

\begin{figure}[t!]
\centering
    \includegraphics[width=\columnwidth]{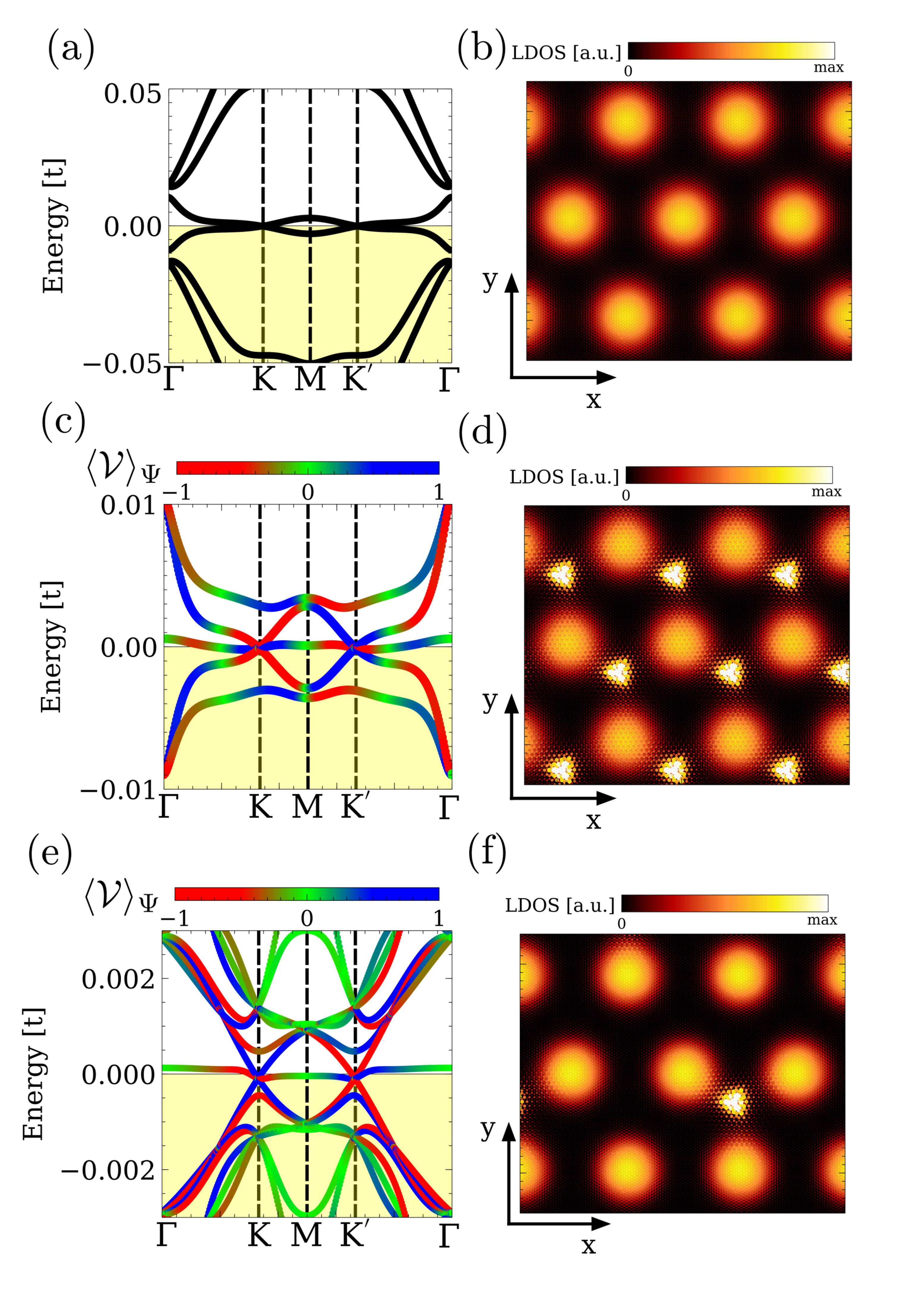}

\caption{
	Band structure
	of pristine twisted bilayer graphene
	at $\alpha = 1.5^\circ$, and (b) local
	density of states  at charge neutrality $E=0$.
	(c) Band structure with a single impurity
	per moire supercell, together
	with the (d) local density of states, showing the
	emergence of zero modes
	around the impurity.
	(e) Band structure for a single vacancy in a
	$2\times 2$ supercell, and (f)
	spatial density of states.
	We took $m_0=22$ and $t_\perp=0.15t$ for all panels.
}
\label{fig3}
\end{figure}

The robustness of the triple point is guaranteed by the effective valley quantum number and the properties of the emergent orbitals at low energies. It can be understood
from the consideration of the band degeneracy of pristine TBG. The
band structure is displayed in Fig. \ref{fig3} (a), with two
double degenerate
bands crossing the Fermi level in the
$\Gamma-K-M$ direction \footnote{The degeneracy in the $\Gamma-M$ direction
is broken already in the pristine system}. 
We can label the eigenstates associated with
the branches $E_{1\bk}$ as $\Psi_{1\bk}$ and $\Psi_{\bar 1\bk}$, and the other
two associated with $E_{2\bk} = -E_{1\bk}$ as $\Psi_{2\bk}$ and $\Psi_{\bar 2
\bk}$. These wave functions have weight in all microscopic degrees of freedom,
layer and sublattice, such that these are expected to have a finite amplitude
at a generic impurity site. Focusing on the first degenerate set, an impurity
introduces a coupling between the eigenstates of pristine TBG such that the
Hamiltonian in the eigenbasis $(\Psi_{1\bk}, \Psi_{\bar 1\bk})$ can be written
as
$
	H_1 (\bk)=
\begin{pmatrix}
		E_{1\bk} + v_1 & \sqrt{v_{\bar1} v _1}  \\
		\sqrt{v_{\bar1} v _1} & E_{1\bk} + v_{\bar1} \\
\end{pmatrix},
	$
where $v_1$ and $v_{\bar1}$ stand for the coupling of the impurity to the
respective state, 
$v_i = w |\Psi_{i\bk}(\br_n)|^2$. This
eigenproblem has solution  $\phi_{1\bk}$ with $E_{1\bk}$ and  $\phi_{\bar 1
\bk}$ with $E_{1\bk}+v_1+ v_{\bar1}$. Note that $\phi_{1\bk}$ remains at its
original energy, indicating that it has zero amplitude at the impurity site and
is therefore blind to its presence, while $\phi_{\bar 1 \bk}$ couples to the
impurity and is shifted in energy. An analogous construction can be made for
states $\Psi_{2\bk}$ and $\Psi_{\bar 2 \bk}$, such that in the appropriate
combination, one of the states decouples from the impurity site, such that $\phi_{1\bk}$ and $\phi_{2\bk}$ give rise to two bands which do not change in  presence of the impurity. 

We now focus on the $K$ point, at which all low lying bands become degenerate.
At this point, we can choose to write the eigenstates in a basis which is
valley polarized. From the numerical analysis, we infer that this basis can, in
fact, be written as $\Psi^\dagger =
(\phi^\dagger_{\bar 1 \bk},
\phi^\dagger_{1\bk},
\phi^\dagger_{2 \bk},
\phi^\dagger_{\bar 2 \bk})$, in terms of the eigenstates discussed in the previous paragraph. In this basis the valley operator yields $\mathcal{V} = \text{diag}(+1,+1,-1,-1)$, and the effective Hamiltonian around the $K$ point is of the form:
\begin{equation}
        H^{eff} (K + \bk)=
        \begin{pmatrix}
                \Lambda_1 & \bar k & 0 & \sqrt{\Lambda_1\Lambda_2} \\
                \bar k^* & 0 & 0 & 0 \\
                0 & 0 & 0 & \bar k^* \\
                \sqrt{\Lambda_1\Lambda_2} & 0 & \bar k & \Lambda_2 \\
        \end{pmatrix}
        \label{heff}
\end{equation}
where $\bar k = \bar v_F(k_x + ik_y)$, $\bar v_F$ is the renormalized Fermi velocity,
and $\Lambda_i = v_i + v_{\bar i}$.
The effective Hamiltonian at $K'$ follows from time reversal symmetry.
In the absence of impurities, $\Lambda_{1,2}= 0$, the doubly degenerate Dirac dispersion is found.
In the presence of an impurity, $\Lambda_{1,2} \ne 0$, the effective
Hamiltonian has eigenvalues $\pm |\bar k|$ and $\frac{1}{2}(\Lambda \pm \sqrt{4
|\bar k|^2 + \Lambda^2})$, where $\Lambda = \Lambda_1+\Lambda_2$. Note that at
the $K$ point ($\bar k =0$) the eigenvalues are $\{0,0,0,\Lambda\}$, making the
triple point explicit and associating the splitting to the magnitude of the
impurity potential as $\Lambda \sim w$. In this generic scenario, states with different valley number are mixed. Note that in the case
$\Lambda_1\ne0$ and $\Lambda_2=0$, a triple point emerges with well defined
valley number, since $[\mathcal{V},H^{eff}] = 0$, as shown in Fig. \ref{fig2}
(b), which happens for impurities in the AB/BA regions, as mapped in Fig.
\ref{fig2} (d). This valley polarization for impurities in the AB/BA regions
can be inferred from the properties of the Wannier wave functions in real space,
as shown, for example, in Ref. \onlinecite{Koshino2018}.

\section{The Vacancy limit}
\label{sec:vac}

We now move on to consider the case of vacancy effects \cite{PhysRevLett.92.225502,Yazyev2010,PhysRevB.75.125408,PhysRevB.77.035427,PhysRevB.77.195428,PhysRevB.96.024403,Sousa2019,PhysRevB.89.245429,PhysRevLett.104.096804}, 
namely, $w\rightarrow
\infty$. This limit is especially attractive because it can be achieved by
adsorbed hydrogen atoms \cite{Brihuega2017}. Recent experiments have shown that it is possible to manipulate \cite{GonzalezHerrero2016} and even automatize the manipulation \cite{Moller2017} of hydrogen atoms with atomic precision by scanning tunneling microscope. A vacancy in monolayer graphene is known to give rise to a zero mode at charge
neutrality, according to Lieb's theorem \cite{Pereira2008}. 
In TBG, this zero
mode will, however, coexist in energy with the nearly flat honeycomb-like bands
of Fig. \ref{fig3} (a) and (b), and therefore it is expected to heavily hybridize with
them. Interestingly, as shown in Fig. \ref{fig3} (c), the hybridization of the
vacancy mode with the honeycomb-like bands lifts one of the Dirac cones and
generates a flat band. Moreover, even though vacancies are expected to create
strong inter-valley scattering, we find that the remaining Dirac cones are
perfectly valley polarized when the vacancy is located at the AB/BA regions,
see Fig. \ref{fig3} (c) and (d) and Fig. \ref{fig3} (e) and (f),
similarly to the weak impurity scenario above.
The weak and strong potential limits can actually be continuously connected by
ramping up the parameter $w$.
In this process, it is observed that the $K$ point
always displays a triple point, two of the bands remain rigid, and the
quadratic band pinned to the Fermi level evolves smoothly towards a flat band.
As the potential is increased, part of the electronic density drifts from the
AA regions to the location of the impurity. This phenomenology is observed to
be independent of the density of vacancies per unit cell, as shown in the
calculation for a $2$x$2$ supercell in Fig. \ref{fig3} (e) and (f)
\footnote{Substantial second neighbor hopping adds a dispersion to the vacancy band, yet preserves the triple point.}. 

\begin{figure}[t!]
\centering
    \includegraphics[width=\columnwidth]{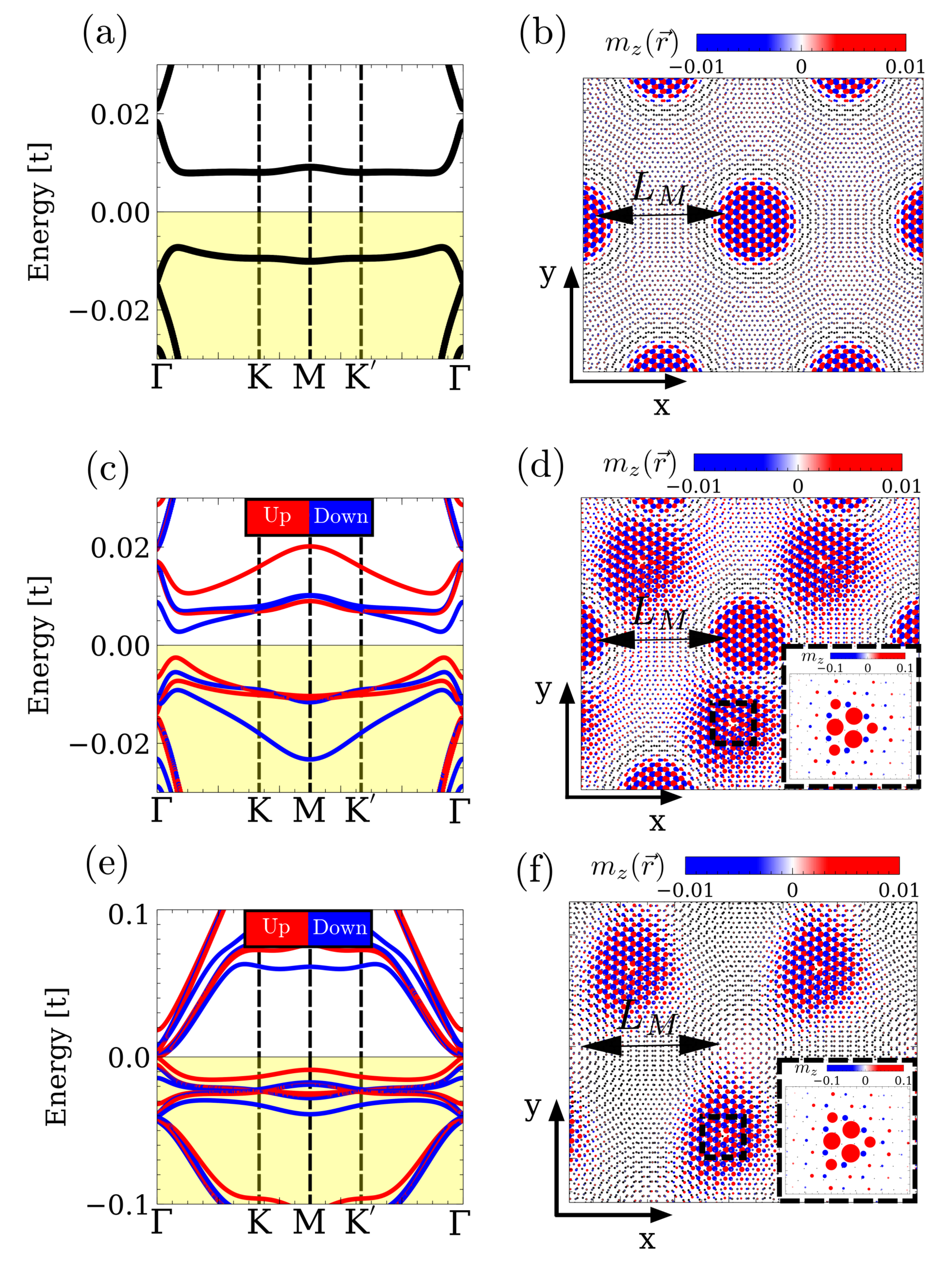}

\caption{
	Self-consistent band structure after including interactions
	for (a) charge neutral pristine twisted bilayer graphene, 
	together with its (b) ground state magnetization.
	Self-consistent band structure of (c) twisted bilayer graphene
	with a single vacancy in the AB region, together with its (d) ground state magnetization.
	(e) The band structure and (f)
	magnetization with a filling of four
	electrons per unit cell with respect to charge
	neutrality.
	Note the different maximum values in (d) and (f) and their inset.
	We took $m_0=11$, $t_\perp=0.3t$ and $U=2t$
	for all panels.
}
\label{fig4}
\end{figure}

\section{Interaction effects}
\label{sec:int}

It is important to note that this discussion relied on a single particle picture. However,
the large density of states associated with these nearly flat bands suggests
that, at low temperatures, a symmetry broken state develops due to interactions \cite{PhysRevLett.120.266402,CaoMott2018,PhysRevB.98.075109,Kang2018,PhysRevLett.121.087001}.
To account for the effect of electronic interactions, we make explicit the spin
degree of freedom in Eq. \ref{hamil}
and introduce an interaction term of the form  $H_U = U
\sum_i	n_{i\uparrow} n_{i\downarrow}$, where $n_{i\sigma}$ counts the number
of electrons with spin $\sigma = \{\uparrow, \downarrow\}$ at site $i$
  \cite{Finocchiaro2017,PhysRevLett.119.107201,PhysRevLett.120.266402,
PhysRevLett.106.236805}. We use a mean field ansatz of the form
$H_U \approx U \sum_i 
	\langle n_{i\uparrow} \rangle n_{i\downarrow} +
	n_{i\uparrow} \langle n_{i\downarrow} \rangle - 
	\langle n_{i\uparrow} \rangle \langle n_{i\downarrow} \rangle$, with the expectation values determined self-consistently, allowing for a local ground state magnetization $m^z_i =\langle n_{i\uparrow} \rangle
	- \langle n_{i\downarrow} \rangle $. For pristine TBG at half filling, Figs. \ref{fig4} (a) and (b) show that interactions drive the system into an insulating state with antiferromagnetic order in the AA regions   \cite{PhysRevLett.119.107201,PhysRevLett.120.266402, PhysRevB.76.184430,PhysRevLett.121.217001}. 
A more interesting scenario takes place in the presence of a vacancy in an AB
region of TBG, in which case electronic interactions create a localized
magnetic moment \cite{PhysRevLett.92.225502,Yazyev2010,PhysRevB.75.125408,PhysRevB.77.035427,PhysRevB.77.195428,PhysRevB.96.024403,Sousa2019,PhysRevB.89.245429}. 
As shown in Figs. \ref{fig4} (c) and (d), the impurity state
is not detrimental to the opening of a gap and the associated antiferromagnetic
ordering of the AA regions, even though the magnetization associated with the impurity states is one order of magnitude larger than the staggered magnetization in the AA regions (see inset).
As a result of the weak antiferromagnetism in the AA regions, doping quenches
the antiferromagnetic order, while the magnetization around the vacancy survives. This can be clearly seen
in Figs. \ref{fig4} (e) and (f), where we consider a doping of four extra electrons per
unit cell, which fills the low energy bands up to their edge.
In fact, this phenomenology holds up to chemical potentials of the order of the
exchange splitting $\sim 30$ meV \cite{GonzalezHerrero2016,PhysRevB.75.125408},
much bigger than the $\sim 8$ meV bandwidth of the nearly flat bands
  \cite{Koshino2018,Po2018,Kang2018,bistritzer2011moire}. 
As a result, at low temperatures, these localized magnetic moments may coexist with other phases
found in the bilayer such as superconducting \cite{cao2018unconventional},
strange metal phases \cite{Cao2019} or
anomalous Hall states \cite{Sharpe2019}.

In the case of larger angles, the Dirac cone states are not expected
to have an electronic instability due to their substantial Fermi velocity.
As a result, the ground state at half filling will be defined by an instability driven only by the impurity bands. These aspects are discussed in detail in Appendix \ref{sec:int3}.

\section{Conclusions}
\label{sec:con}

To summarize, we have established that impurities in small-angle twisted bilayer graphene give
rise to robust triple point fermions at the charge neutrality point,
independently of the impurity potential and 
its location.
We have shown that the
triple point modes can be valley polarized for defects located in the AB and BA
regions, providing a route to engineer triple-point fermions with an additional
quantized degree of freedom. 
In the presence of interactions, 
the triple points can be lifted by the development of magnetic order, introducing the possibility
of engineering correlated states of triple point fermions in 
twisted bilayer graphene.
Our results put forward a new mechanism to generate triple-points in graphene
systems, providing a starting point to study their intrinsic properties and
interplay with additional emergent states in twisted bilayers.

\acknowledgments
We would like to thank C. Timm, D. F. Agterberg, 
P. Brydon, H. Menke, T. M. R. Wolf,
O. Zilberberg, G. Blatter,
W. Chen, B. Amorim, F. Guinea, and
E. V. Castro
for helpful discussions.
AR acknowledges financial support from FAPESP JP project (2018/04955-9) and fellowship
(2018/18287-8), and Fundunesp/Simons Foundation (2338-2014 CCP).  AR is also
grateful for the hospitality of the Pauli Centre of ETH Zurich.
J.L.L acknowledges financial support from the ETH
Fellowship program
and from the
JSPS Core-to-Core program ``Oxide Superspin" international network.

\appendix

\section{Triple points for larger twisting angles}
\label{sec:largealpha}

Above we focused on a small
twisting angle, $\alpha \approx1.5^\circ$, in which case the
Dirac cones show a highly reduced Fermi velocity.
In this scenario, in the presence of a vacancy, the system
presented two types of localized modes, i.e.,  the vacancy mode and the
nearly flat honeycomb band, forming a triple point at $K$.

\begin{figure}[t!]
\includegraphics[width=\columnwidth]{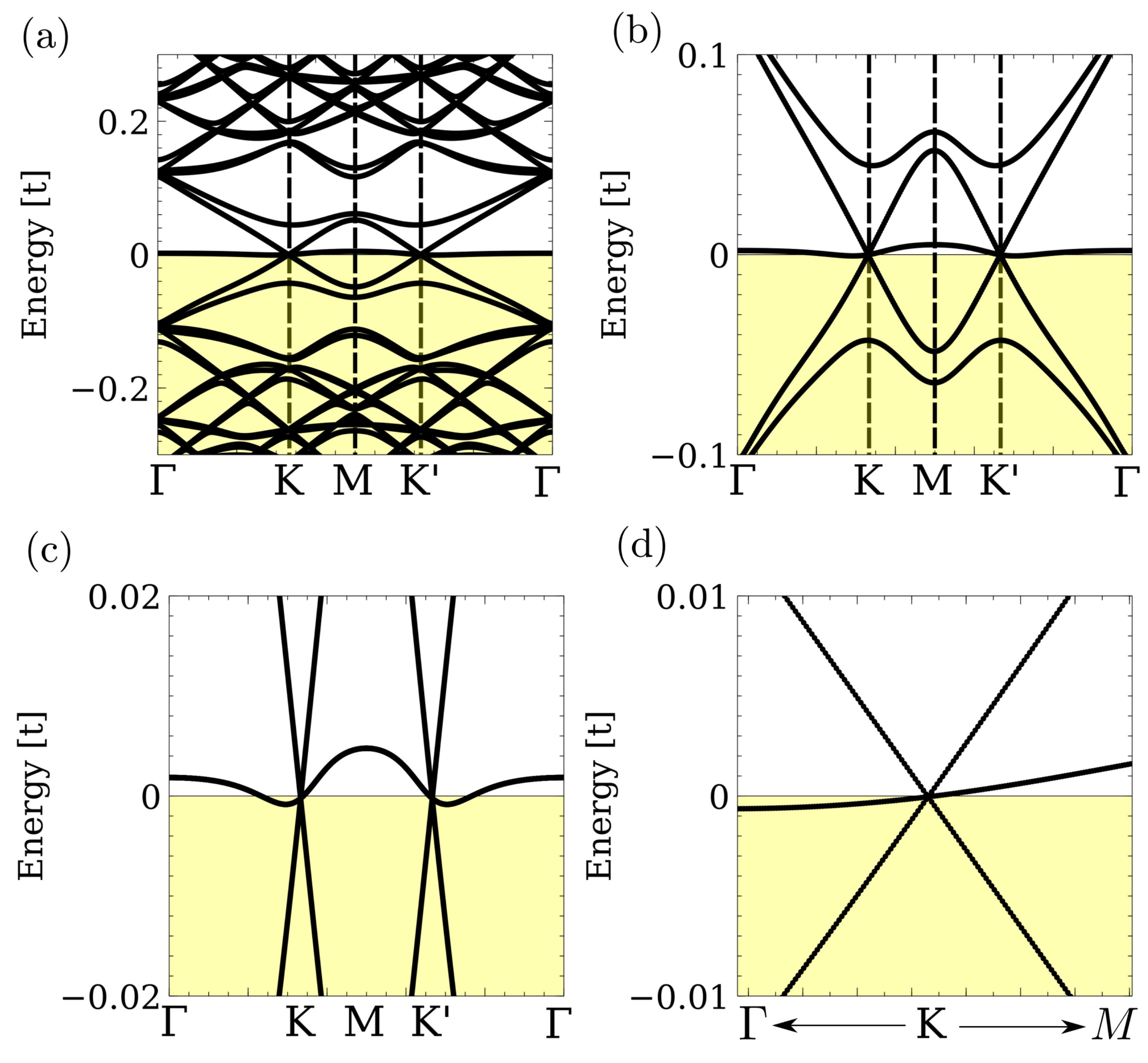}
\caption{ (a) Band structure of twisted bilayer graphene
        with a vacancy in the AB region, for a a twisting
        angle of $\alpha \approx 3^\circ$ between the two layers.
      (b), (c) A zoom closer to the Fermi energy of (a) the band structure,
        and (d) a zoom close to the $K$ point highlighting
        the triple crossing. We took $m_0=11$ and $t_\perp =0.12t$.}
\label{SMfig1}
\end{figure}

\begin{figure}[t!]
    \includegraphics[width=\columnwidth]{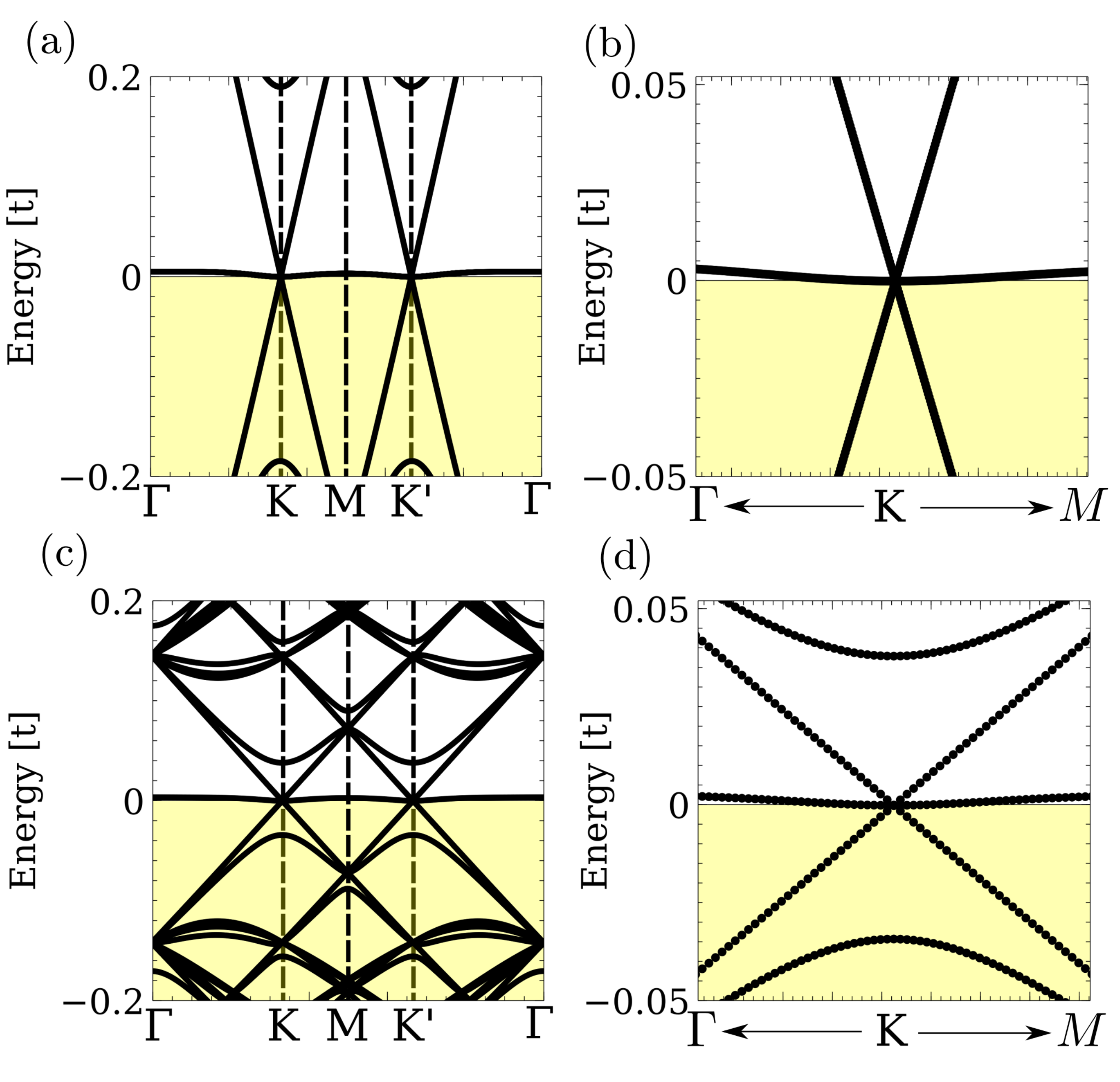}
	\caption{ (a) Band structure for
	twisted bilayer graphene at an angle $\alpha\approx 9.5^\circ$
	with one vacancy per moire unit cell, showing the existence of
	triple point crossing even at large angles (zoom in (b)).
	(c) Band structure for $\alpha\approx 9.5^\circ$,
	with one vacancy for a $4x4$ moire supercell, showing the
	persistence of triple point crossings (zoom in (d)).
}
\label{SMfig3}
\end{figure}

The emergence of a triple point is not a unique feature of small angles. Here we show in Fig. \ref{SMfig1} and \ref{SMfig3} the band structure for angle $\alpha \approx 3^\circ$ and $\alpha \approx 9.5^\circ$, respectively. Both structures show the emergence of a flat band, with a triple point crossing. In comparison with the case $\alpha\approx 1.5^\circ$, the Dirac cones for larger angles show a much higher Fermi velocity, a feature that does not affect the presence of the triple point. In order to show the robustness of the triple point within the tight-binding calculation, we also show figures which zoom in energy and around the $K$ point for both angles.

\section{Interaction effects for $\alpha \approx 3^\circ$}
\label{sec:int3}

In the case of larger angles shown above, the Dirac cone states are not expected
to have an electronic instability due to their substantial Fermi velocity.
As a result, the ground state at half filling will be defined by an instability driven only by the impurity bands, 
different than the one presented for $\alpha\approx 1.5^\circ$ in the main text, where both flat Dirac cones and impurity bands contribute to the instability.

\begin{figure}[t]
\centering
    \includegraphics[width=\columnwidth]{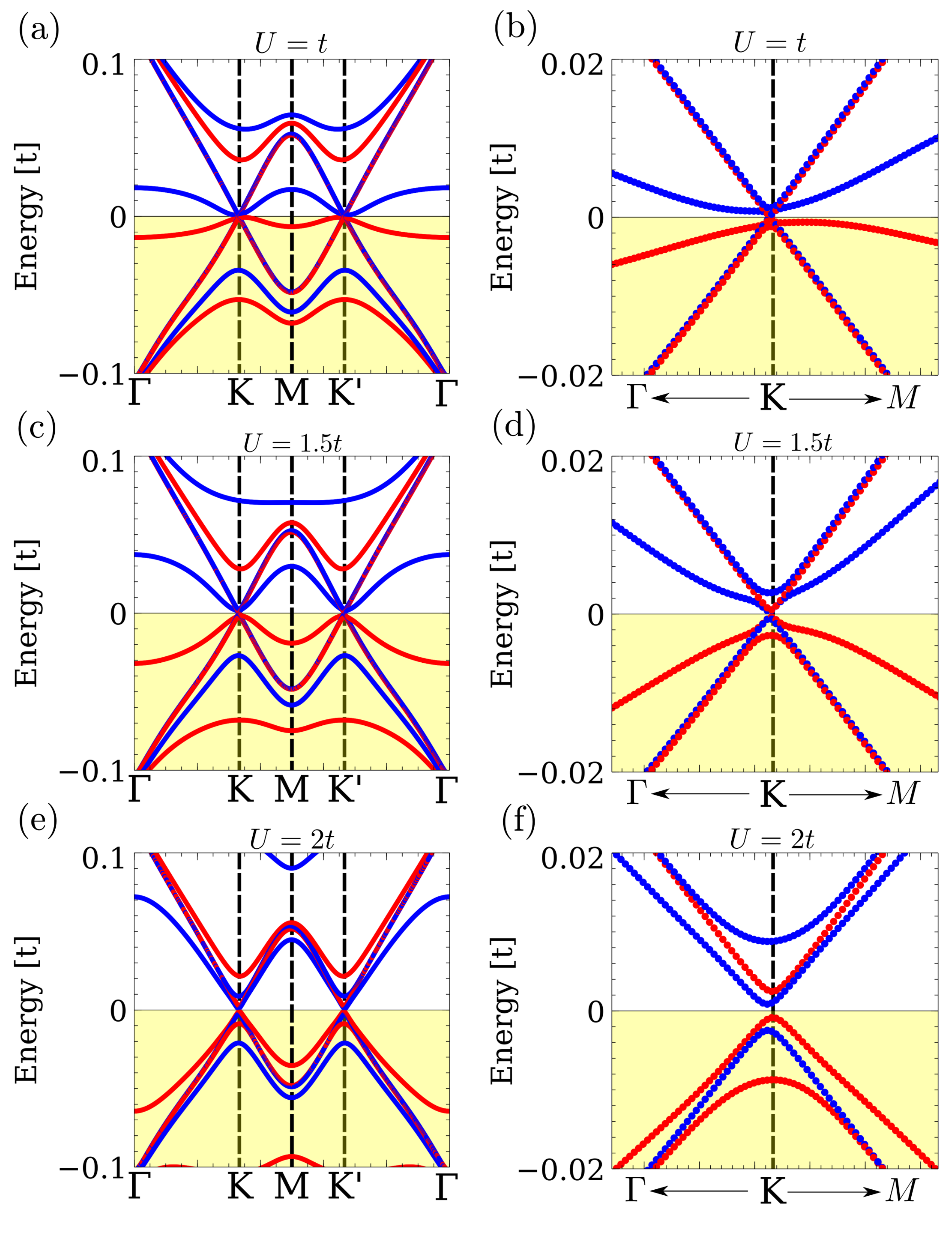}

        \caption{Selfconsistent band structure
for twisted bilayer graphene with a vacancy in the AB region,
        for an angle $\alpha \approx 3^\circ$, for (a) $U=t$,
      (c)  $U=1.5t$  and
(e) $U=2t$ , with (b), (d), and (f) the respective zooms around the original Dirac point. The higher interaction, the larger the exchange splitting of the vacancy bands. The color red/blue denotes the expectation value of $S_z=\pm 1$. We took $m_0=11$ and $t_\perp=0.12t$ for all panels.
}
\label{SMfig2}
\end{figure}

Here we focus on the vacancy case for $\alpha\approx 3^\circ$, which presents
a nearly flat band coexisting with a Dirac cone.
For larger angles the electronic instability only takes
place in the vacancy flat band, giving rise to a net
magnetic moment of $ 1\mu_B$ per unit cell
as shown in Fig. \ref{SMfig2}. In particular,
we show in the different panels
of Fig. \ref{SMfig2} that the electronic structure of the
system is qualitatively similar
for the different values of $U$ ranging
from $U=t$ to $U=2t$. It is particularly evident that the vacancy states are highly polarized, while the Dirac cones are split and shifted, what can only  be clearly observed from Fig. \ref{SMfig2} (f).
\newpage

\bibliographystyle{apsrev4-1}
\bibliography{paper}{}

\end{document}